\documentclass[runningheads]{llncs}

\usepackage[T1]{fontenc}

\let\llncssubparagraph\subparagraph
\let\subparagraph\paragraph
\usepackage{titlesec}
\let\subparagraph\llncssubparagraph

\usepackage{graphicx}
\usepackage{tikz}
\usetikzlibrary{trees}
\usepackage{multirow}
\usepackage{siunitx}
\usepackage{enumitem}
\usepackage{url}
\usepackage{xcolor}
\usepackage{tabularx}
\usepackage{booktabs}
\newcolumntype{Y}{>{\centering\arraybackslash}X}

\begin{document}
\title{Relationships are Complicated! An Analysis of Relationships Between Datasets on the Web}

\titlerunning{An Analysis of Relationships Between Datasets on the Web}

\author{Kate Lin \and
Tarfah Alrashed \and
Natasha Noy}
\authorrunning{K. Lin et al.}
\institute{Google Research, Google, USA \\
\email{\{kateslin,tarfah,noy\}@google.com}}

\maketitle              

\begin{abstract}
The Web today has millions of datasets, and the number of datasets continues to grow at a rapid pace. These datasets are not standalone entities; rather, they are intricately connected through complex relationships. Semantic relationships between datasets provide critical insights for research and decision-making processes. In this paper, we study dataset relationships from the perspective of users who discover, use, and share datasets on the Web: what relationships are important for different tasks? What contextual information might users want to know? We first present a comprehensive taxonomy of relationships between datasets on the Web and map these relationships to user tasks performed during dataset discovery. We develop a series of methods to identify these relationships and compare their performance on a large corpus of datasets generated from Web pages with \path{schema.org} markup. We demonstrate that machine-learning based methods that use dataset metadata achieve multi-class classification accuracy of 90\%. 
Finally, we highlight gaps in available semantic markup for datasets and discuss how incorporating comprehensive semantics can facilitate the identification of dataset relationships. By providing a comprehensive overview of dataset relationships at scale, this paper sets a benchmark for future research.


\end{abstract}
\section{Introduction}

As the world becomes increasingly data-driven, researchers rely on open data to answer scientific questions and to understand complex phenomena \cite{zuiderwijk2020drives}. This reliance on data has led to dataset publication becoming the norm in many scientific disciplines \cite{tedersoo2021data}. Unlike scientific publications, however, datasets are not static and standalone entities: dataset providers publish new versions of datasets as the data evolves, researchers produce new datasets by combining existing datasets, and meaningful subsets of large datasets may gain a life of their own. When a user chooses a dataset for her work, these distinctions become critical. For instance, when reproducing results from a publication, we must identify which specific dataset snapshot the authors used. When evaluating the trustworthiness of a dataset available on multiple platforms, users may want to choose the repository that they trust the most. If a scientist wants to compare slices of a large dataset, she wants to ascertain that these slices come from the same snapshot of the larger dataset. Therefore, understanding the semantics of relationships between datasets can be just as important as understanding other metadata about them.

How can we identify relationships between datasets? The most straightforward method is to look at information provided by dataset publishers. The industry standard for describing semantic metadata for any Web content (including datasets) is through structured markup in \path{schema.org}~\cite{guha-schema}. Standards like \path{schema.org} and W3C DCAT~\cite{DCAT} provide means to identify pages containing datasets as well as the semantics of relationships between them. In addition to these Web standards, approaches such as datasheets~\cite{datasheets} provide mechanisms to describe dataset origins, biases, and recommended usage. These methods of providing additional context for datasets enable publishers to link versions of the same dataset to one another, link a dataset in one repository to the original dataset in another repository, or to declare that one dataset is based on another. However, semantic markup is often unreliable and incomplete,~\cite{alrashed-datasetness,meusel-schema,hogan2010weaving} and only a small fraction of dataset metadata on the Web contains values for properties that link them to other datasets~\cite{benjelloun-2020}. Furthermore, dataset authors often update or restructure datasets without providing notice or documentation~\cite{Kery-2019,Zhang-2020,Umbrich2010TowardsDD}. Finally, current markup frameworks do not fully capture the variety and nuance of dataset relationships. 

\begin{figure}
\begin{tikzpicture}[
    level 1/.style = {sibling distance = 3.5cm},
    level 2/.style = {sibling distance = 2.5cm},
    level 3/.style = {sibling distance = 3cm, text width=2cm},
    edge from parent fork down,
    align=center,
    text width=2cm]
\node [text width=\linewidth]{Aquarius Official Release Level 3 Ancillary Reynolds Sea Surface Temperature Standard Mapped Image}
    child {node (left4 node){Annual V4}
            child {node {Ascending Annual V4}
                child {node {Ascending Annual V5}}
            }
            child {node (left5 node){Annual V5}}
    edge from parent}
    child {node (middle4 node){Monthly V4}
            child {node (middle5 node){Monthly V5}
                 child {node {Descending Monthly V5}}
            }
    edge from parent}
    child {node (right4 node){Daily V4}
            child {node (right5 node){Daily V5}
                child {node {7-Day V5}}
            }
            child {node {28-Day V4}
                 child {node {28-Day V5}}
            }
    };
  \draw[dashed,-] (left4 node) -- (middle4 node);
  \draw[dashed,-] (middle4 node) -- (right4 node);
  \draw[dashed,-] (left5 node) -- (middle5 node);
  \draw[dashed,-] (middle5 node) -- (right5 node);
\end{tikzpicture}
\caption[engine features]{The \textit{``Aquarius Official Release Level 3 Ancillary Reynolds Sea Surface Temperature Standard Mapped Image''} dataset has annual, monthly, and daily variants with multiple versions. Variants are derived from each other and can have different replicas (e.g., \textit{``Annual V4''} on three sites) and reconfigurations (ascending, descending).
\protect\footnotemark[2] }
\label{fig:example}

\end{figure}
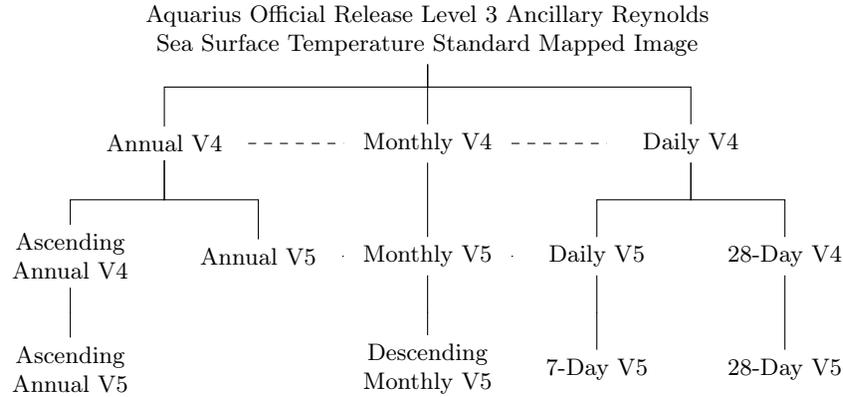

To illustrate the richness of dataset relationships, consider the collection of datasets provided by the United States Census Bureau. This collection captures a wide variety of measures (e.g., income data) over many decades at various levels of granularity, such as national, statewide, county-wide, and so on. On the Web, one can find various slices of this large dataset that may be relevant in a specific context: for example, there may be a dataset containing income data for California in 2008 or a dataset containing income data for the entire US in the same year. Each of these datasets is a subset of the larger 2008 Census dataset, but researchers may publish them in different Web sites and contexts.  Figure~\ref{fig:example} presents another example---a collection of datasets published by the US National Oceanic and Atmospheric Administration (NOAA). We have found more than 140 forms of this dataset on the Web, including annual, monthly, and daily sets. Many variations have multiple versions, and some of these datasets are subsets of larger ones. Additionally, many of the datasets in this collection are replicated across various Web sites. Critically, the information that helps us understand these relationships often must come from \emph{metadata}: the data itself may not have enough context for us to understand the provenance or coverage of the datasets.

In this paper, we explore the relationships between datasets from the perspective of users who want to discover and analyze datasets. Rather than define these relationships in an abstract way, we take a user-centric view and ground the relationships in user tasks performed during dataset discovery. We analyze a large dataset corpus, generated from dataset pages on the Web with \path{schema.org/Dataset} markup, to identify these relationships between datasets. Our evaluation focuses specifically on provenance-based relationships, which we can infer from metadata. There is a large body of work (Section~\ref{sec:related-work}) that infers relationships from data. Our focus on metadata and specifically on provenance-based relationships complements related work.

Specifically, we make the following contributions in this paper:
\begin{itemize}
    \item We define a taxonomy of relationships between datasets on the Web and ground it in essential user tasks that rely on understanding these relationships (Sections~\ref{sec:user-tasks} and~\ref{sec:dataset-relationships}). To our knowledge, this taxonomy is the most comprehensive in the literature to date.
     \item We propose and compare several methods for identifying dataset relationships (Sections~\ref{sec:methods}). We show that machine-learning methods that use dataset metadata achieve a multi-class classification accuracy of 90\%, outperforming \path{schema.org} and heuristics-based methods (Section~\ref{sec:evaluation}).
     \item We analyze the prevalence of provenance-based relationships in a corpus of 2.7 million datasets on the Web. We found that 20\% of datasets have at least one relationship with another dataset (Section~\ref{sec:evaluation}).
     \item We present recommendations for enhancing dataset metadata, facilitating the discovery of more relationships between datasets
      (Section~\ref{sec:discussion}).
     \item We publish a collection of 2.7 million dataset pages with their basic metadata, along with the connections and interrelationships among these datasets.\footnote{ \url{https://figshare.com/articles/dataset/Metadata_for_Datasets_and_Relationships/22790810}}
\end{itemize}

\footnotetext[2]{See, for example, dozens of versions of this dataset in Google Dataset Search: \url{https://bit.ly/3IrdEb2}}

\section{Related Work}
\label{sec:related-work}

Our research focuses on relationships between datasets on the Web. In this section, we examine approaches to tackling this topic, including exploring dataset provenance, understanding dataset evolution, identifying conceptually similar or joinable datasets, and linking datasets to scientific publications.

Research reproducibility benefits from understanding dataset provenance and relationships between datasets. Herschel et al. \cite{herschel2017survey} proposed user-focused methods for capturing and analyzing provenance, emphasizing relationship presentation. Klump et al. \cite{Klump-2021} introduced a contextual framework and versioning principles surpassing simple revisions. Rauber et al. \cite{rauber-2016} identified subsets of large datasets supporting research findings, while Silvello \cite{silvello2015methodology} detailed citing linked open data subsets. Rauber et al. \cite{rauber-2015} stressed the importance of exact version citation for research reproducibility.

Semantic vocabularies like \path{schema.org}\cite{guha-schema}, DCAT\cite{DCAT}, and VoID~\cite{VoID} enable dataset providers to specify relationships between datasets (e.g., \path{schema.org/isBasedOn} for derivations, \path{schema.org/sameAs} and \texttt{void:Linkset} for replicas, DCAT's \texttt{dct:isVersionOf} for versioning). However, these standards lack a comprehensive vocabulary for diverse dataset relationships. The PROV Data Model~\cite{PROV} offers a standardized framework for describing provenance information between entities, including \texttt{wasDerivedFrom} for dataset derivation. Indeed the PROV notion of derivation can form the basis for some of the relationships that we discuss later. However, because datasets and their relationships are not the focus of PROV, it does not capture many of the dataset-specific relationships, such as subsets, slices, or datasets that can be integrated.

Research in data evolution addresses challenges of detecting, tracking, and explaining dataset changes over time. Roussakis et al. \cite{roussakis-2015} introduced a flexible framework for dynamic dataset analysis, offering various granularity levels and rich visualizations. Umbrich et al. \cite{Umbrich2010TowardsDD} quantified change frequency in linked open data (LOD) for improved understanding of dataset dynamics. Shraga and Miller \cite{shraga2023explaining} proposed a semantic data versioning method using explanations to aid users in comprehending dataset changes. 

Other LOD research focuses on recommending interconnected datasets. Lopes et al.~\cite{lopes-2014} proposed collaborative filtering and content-based methods using dataset metadata, both validated through real-world evaluation. Leme et al.~\cite{leme-2013} suggested using core datasets for interlinking recommendations, demonstrating effectiveness on a large-scale dataset. Ellefi et al.~\cite{ellefi-2016} introduced an intensional approach based on conceptual similarity. The LOD research focuses on linking entities within datasets~\cite{Faerber-2022,volz2009,10.1145/3591366.3591378,10.1016/j.websem.2018.12.003}, complementary to our focus on complete datasets rather than individual datapoints.

In addition to finding conceptually similar datasets, researchers emphasize identifying joinable datasets to enrich information without foreign keys \cite{10.1145/3555041.3589409}. Zhu et al. introduced Josie to find joinable tables using overlap set similarity search \cite{zhu-2019}. Dong et al. developed a clustering-based method for grouping joinable tables \cite{dong2021efficient}. Recent work includes DeepJoin, a deep learning model for efficient joinable table discovery \cite{dong2023deepjoinjoinabletablediscovery}. These efforts streamline dataset discovery for data analysts.

Efforts to link datasets with scientific publications are significant. The Research Graph dataset~\cite{aryani2018research} consolidates publication, dataset, and software details into a unified knowledge graph, simplifying research output discovery. Ayush et al.~\cite{singhal2017research} applied natural language processing to extract data-related information from publications and match it to online datasets. Google Dataset Search~\cite{datasetsearch} connects datasets with scholarly articles referencing them, enhancing scientific discovery by simplifying dataset finding for researchers.

The highlighted research emphasizes the importance of dataset relationships in scientific research. Our aim is to offer a broader framework for future research by identifying more relationships, evaluating their impact, and thus advancing the development of better dataset management tools and techniques.

\section{Grounding Dataset Relationships in User Tasks}
\label{sec:user-tasks}

The data ecosystem relies on a continuous cycle of discovery, analysis, and sharing, necessitating a nuanced understanding of dataset relationships. To ground our taxonomy of dataset relationship, we begin by defining the tasks users undertake during data discovery and sharing.

\textbf{Finding Datasets:}
The proliferation of data on the Web has complicated dataset discovery through traditional search engines~\cite{Koesten-2017}. Users face challenges in sorting through vast amounts of data and navigating diverse search criteria, especially when their intent varies. For instance, reproducing an experiment from a paper demands sorting through dataset versions, formats, and sources~\cite{Klump-2021}, while augmenting a dataset requires finding compatible data in structure, schema, and topic. Understanding dataset relationships facilitates efficient and accurate dataset selection, aiding in reliable data-driven decision-making.

\textbf{Evaluating Dataset Trustworthiness:}
Evaluating whether to use a dataset involves an assessment of dataset trustworthiness \cite{herschel2017survey,Gregory-2020}. Unlike research publications, datasets published on the Web rarely undergo peer review. As a result, users must rely on dataset attributes and metadata as proxies for dataset trustworthiness. User-experience studies reveal users weigh data provider, format, prior usage, and update frequency \cite{Gregory-2020}. Hence, comprehending dataset provenance, identifying citations, and locating reliable sources aid in evaluation.

\textbf{Citing and Referencing Datasets:}
Noy et al. \cite{noy-2022} note that well-described datasets drive new research. Citing datasets like papers encourages better data collection and curation \cite{herschel2017survey}. Proper citation requires persistent identifiers, metadata, and accurate provenance descriptions, including version number, source, and whether it is a subset of another dataset \cite{Klump-2021}. Proper dataset identification promotes transparency and collaboration, enhancing research quality.

\textbf{Curating Datasets:}
Dataset curation involves collecting, organizing, and maintaining datasets from diverse sources to ensure availability for users. The goal is to create high-quality datasets beneficial for researchers, developers, and users. Curators must understand a dataset's relationships with others, including versions, replicas, and usage in research or projects \cite{datasheets}.

While the list of user tasks in this section is by no means exhaustive, it demonstrates the need for improving our understanding of the relationships between datasets, capturing these relationships explicitly, and using these relationships to improve researchers' experiences with data.

\section{Defining Dataset Relationships}
\label{sec:dataset-relationships}

We base our categorization of dataset relationships on the analysis of user tasks in dataset discovery (Section~\ref{sec:user-tasks}), prior research (Section~\ref{sec:related-work}), and observations from analyzing a large corpus of datasets on the Web. Specifically, we collected a corpus of datasets by relying on \path{schema.org/Dataset} markup (Section~\ref{sec:data}).

We broadly group dataset relationships on the Web into provenance-based and non-provenance-based relationships. \emph{Provenance-based relationships} are relationships between the datasets that share a common original dataset, such as being derived or modified versions of the same original dataset.
\emph{Non-provenance-based relationships} involve connections between datasets based on content, topic, or task rather than their origin. 
For each relationship, we highlight which of the tasks from Section~\ref{sec:user-tasks} it is particularly useful for.

\subsection{Provenance-based Relationships}

We define a dataset on the web $D = (P, O, S, W)$ as consisting of a set of data points $P$, origin of dataset $O$ (i.e., primary data-collection process), schema $S$, and a web site that hosts the dataset $W$. Note that in the context of dataset discovery, we may not have complete information about a given dataset and, in particular, may not know origin or schema.

We define schema $S$ for the most common dataset types in our corpus. Relational (tabular) datasets are characterized by a header row with column names, data types, and constraints such as primary keys. Document datasets, such as PDFs, are defined by collections of key-value pairs, arrays, or nested documents. Structured datasets, like JSON, are defined by the structure and properties of elements, including data types, relationships, and constraints. RDF data is defined by RDF Schema.

\subsubsection{Replica:} 
Datasets $D_1 = (P_1, O_1, S_1, W_1)$ and $D_2 =  (P_2, O_2, S_2, W_2)$ are \emph{replicas} of each other iff their underlying data and origin are identical, but they are hosted on different sites: $P_1 = P_2$, $O_1=O_2$, $S_1=S_2$, and $W_1 \neq W_2$.

In today's Web data ecosystem, a common pattern is one repository aggregating datasets from multiple other repositories. For instance, the European data portal (europa.eu) offers access to datasets from member state data portals. In the United States, an open data site for a state government might include datasets from county data repositories, which in turn may include datasets from individual towns. Therefore, datasets from local governments (e.g., county) are \emph{replicated} in the state repository. Identifying and grouping replicas of datasets in a dataset-discovery context helps users easily locate datasets and provides choices of sources. It also enables users to obtain data from their most trusted source when available from multiple sources. For example, a user may trust their local government site more and opt to retrieve the dataset from there.

\subsubsection{Version and Revision:}
Datasets $D_1 = (P_1, O_1, S_1, W_1)$ and $D_2 =  (P_2, O_2, S_2,\\ W_2)$ are \emph{versions} of each other iff 
$P_1 \approx P_2$, $O_1 = O_2$, $S_1 \approx S_2$. $W_1$ may or may not be the same as $W_2$. This relationship captures evolution of a dataset over time, where changes between subsequent versions are usually relatively small to the size of the dataset.

Published datasets resemble software more than research publications, as they continue evolving after release. Evolution can range from error corrections and data adjustments to continuous updates with new observations. The Research Data Alliance Data Versioning Working Group stipulates that any alteration to a dataset forms a new version that authors must identify, encompassing minor changes such as data additions/removals \cite{rauber-2015}. However, authors often label only significant checkpoints as new versions, thus, we refer to stable, labeled releases as versions. Minor, unlabeled changes are revisions. For instance, if a dataset covers sales data from January to November and is updated with December data and give it a new label, they create a new version of a dataset. Converting Fahrenheit to Celsius in a weather dataset is typically a revision, not a new version. Identifying all dataset versions is crucial for research reproducibility, data quality assessment, and maintaining transparency and proper attribution when citing the data source.

\subsubsection{Subset:}
A dataset  $D' = (P', O', S', W')$ is a \emph{subset (or slice)} of  $D =  (P, O, S, W)$ iff $P' \subset P$, $O'=O$, $S' \subset S$. $W'$ may or may not be the same as $W$. There is usually an extraction function $F(x)$ that determines which data points from $P$ are in $P'$: $P' = \{x \in P | F(x) = true\}$

A dataset subset is a smaller, more focused set of data extracted from a larger dataset, published independently. The subset typically contains data selected based on specific criteria, like time period, geographical region, or variables. For example, a dataset containing weather information for a country may have subsets for specific regions or time periods. Researchers using a subset of a dataset in their work benefit from transparency, accuracy, and appropriate attribution.

Note that there is a complementary \textbf{superset} relationship. For simplicity, we refer only to the subset relationship in the paper.

\subsubsection{Derivation:}
A dataset  $D' = (P', O', S', W')$ is a \emph{derivation} from a collection of datasets $\Delta = \{D_1, \dots, D_n\}$ iff there exists a derivation function $M(x_1, \dots , x_m)$ that transforms, combines, or otherwise manipulates data points for datasets in $\Delta$. Thus, $O'$ is different from $O_1, ..., O_n$. Schemas and web sites may or may not be the same.

A healthy data ecosystem enables users to build new datasets from published ones. A dataset can be derived from one or more datasets as a result of transforming, aggregating, or otherwise manipulating existing datasets. Examples of derived datasets include summaries, aggregations of multiple datasets, and variables created by combining or transforming existing variables. Understanding which datasets served as input for a given dataset can help users evaluate trustworthiness of datasets and understand whether a given dataset has properties they are looking for. For dataset curation in particular, it is usually not sufficient to specify a dataset is derived from another dataset; rather, dataset authors must provide details on changes and modifications to ensure users understand implications and could reproduce the dataset\footnote{Technically, we can consider a subset to be a derived dataset. We distinguish between the two relationships in our taxonomy; for derived datasets, their authors applied some processing or analysis to the data, while for subsets they simply selected the data from an existing dataset.}.

\subsubsection{Variant:}
Datasets $D_1 = (P_1, O_1, S_1, W_1)$ and $D_2 =  (P_2, O_2, S_2, W_2)$ are \emph{variants} of each other iff $P_1 \cap P2 \approx \emptyset$, $O_1 = O_2$, $S_1 = S_2$. $W_1$ may or may not be the same as $W_2$. 

Consider two weather datasets covering different regions of the country and different years; the two datasets use the same schema and were collected in the same way. Teams in these regions may have collected these datasets independently or may have generated them by creating subsets of a larger dataset. The variant relationship captures the link between these two ``sibling'' datasets. Formally, two datasets are variants of each other if they share the schema, origin, and collection methods but differ in coverage along some dimension. This dimension is often temporal or spatial: the same statistics may be collected and published annually over multiple years, for example. These annual datasets would be variants of each other. Identifying dataset variants allows users to compare and analyze datasets to identify patterns and trends that may be obscured if we look only at one dataset in isolation. By comparing variants of datasets, users gain a more comprehensive understanding of the phenomena that the data represents and make more informed decisions.

\subsection{Non-provenance-based Relationships}
With the wealth of datasets on the Web, users can gain useful insights from serendipitous relationships between datasets. These post-hoc relationships can be based on metadata, dataset usage, or similarity in content. Many of these relationships are context-dependent:  a specific dimension makes sense in the context of a specific user task. We define several such relationships in this section. This list is not exhaustive as we cannot predict all possible uses of datasets.

\subsubsection{Topically Similar:} Datasets $D_1$ and $D_2$ are \emph{topically similar} iff their topical similarity score exceeds a defined threshold $\theta$: $Sim(D_1, D_2) \ge \theta$, where $Sim$ represents a similarity function that yields a value within the range of 0 (indicating no similarity) to 1 (reflecting perfect similarity). The choice of threshold $\theta$ and similarity function $Sim$ relies on the context and desired similarity level. The function $Sim(D_1, D_2)$ can be implemented using methods like cosine similarity, Jaccard index, or others, depending on the nature of the datasets at hand.

Topically similar datasets cover the same subject or capture similar topics along the dimensions relevant to the user context. For instance, datasets covering ocean temperature and salinity can be topically similar when understanding the effects of climate on the oceans. In a different context, a dataset of stock prices might be compared to a benchmark index to assess performance; thus these two datasets would be topically similar.

\subsubsection{Task-similar:} 
Datasets $D_1$ and $D_2$ are \emph{task-similar} if $Sim(T(D_1), T(D_2)) \ge \theta$, where $T(D_1)$ and $T(D_2)$ are the tasks for which the two datasets have been designed, and $Sim$ is a similarity function that yields values between 0 (indicating no similarity) and 1 (reflecting perfect similarity). The selection of the threshold $\theta$ and similarity function $Sim$ depends on the specific context and the degree of similarity one aims to capture. 

Dataset metadata may include not only intrinsic properties of a dataset but also a collection of tasks that a dataset may be best suited for or that dataset creators had in mind. Task-similar datasets share similarities in the tasks or problems they are used for. Datasets created for similar tasks allow for comparison and benchmarking of different algorithms or models, aiding in the evaluation and selection of the best model for a given task. For example, Human3.6M and KITTI  datasets are each used for video prediction, yet their subjects vary drastically: humans versus cars.

\subsubsection{Integratable:} 
Datasets $D_1$ and $D_2$ are \emph{integratable} if they share schema or content enabling the integration. Integratable datasets, $D_1$ and $D_2$, are \textit{joinable} when their attribute sets have a non-empty intersection, serving as common attributes or foreign keys: $A(D_1) \cap A(D_2) \neq \emptyset$, where $A$ represents the attribute set (data fields or columns) in the dataset. Integratable datasets, $D_1$ and $D_2$, are \textit{unionable} when their attribute sets have a non-empty intersection $A(D_1) \cap A(D_2) \neq \emptyset$, and they share similar schemas $S_1 \approx S_2$, and the overlap between their  data points is insignificant: $P_1 \cap P_2 \approx \emptyset$.

Integratable datasets share schema or content, allowing their combination. Datasets are \textit{joinable} if they share common attributes or foreign keys, like traffic patterns and accident reports linked by location and time. Datasets are \textit{unionable} if they capture similar data about complementary concepts, such as weather patterns in different cities. Unionable datasets differ from variants as they do not have the same schema or collection methodology.

\subsection{Discussion}

A dataset can have \emph{multiple relationships} with other datasets. For instance, a national dataset of education statistics with state-level information can be a source for multiple state-specific datasets. The state-specific datasets are \emph{variants} of each other, but all the state-specific datasets are the \emph{subsets} of the national dataset. Two datasets can have multiple relationships with each other. For example, a Monthly dataset and a Daily dataset in Figure~\ref{fig:example} are \emph{variants} of each other. However, if the Monthly dataset was created by aggregating the Daily datasets, it is also \emph{derived from} the Daily datasets.
  
Certain relationships are \emph{bidirectional}, while others are \emph{directional}. For instance, if datasets $D_1$ and $D_2$ are replicas, then $D_1$ is a replica of $D_2$ and $D_2$ is also a replica of $D_1$ (bidirectional $D_1 \leftrightarrow D_2$). However, if $D_1$ is a subset of or derived from $D_2$, it implies that $D_1 \rightarrow D_2$ is true, but not $D_2 \rightarrow D_1$.

Dataset providers can explicitly capture some relationships that we identify in this section. Specifically, \path{schema.org} supports  two of these relationships: \emph{replica} (\path{schema.org/sameAs}) and \emph{derivation} (\path{schema.org/isBasedOn}). For other provenance-based relationships, we can rely on analyzing metadata (e.g., \emph{versions}, \emph{variants}) or the  data itself (e.g.,  \emph{integratable} datasets). \emph{Topical and task similarity} relationships depend on the user context. Because \path{schema.org} metadata is not always reliable \cite{alrashed-datasetness,meusel-schema,hogan2010weaving}, combining it with metadata analysis helps identify relationships.

In our empirical analysis, we prioritize metadata, deferring relationships reliant on data or user context for future exploration. Moreover, extensive research exists on deriving dataset relationships directly from the data itself \cite{zhu-2019,dong2021efficient,khatiwada-2022}.

\section{Empirical Analysis Methods}
\label{sec:methods}
Earlier research has shown that  \path{schema.org} metadata is not always reliable \cite{alrashed-datasetness,meusel-schema,hogan2010weaving}. Furthermore, the markup for dataset relationships is extremely incomplete~\cite{datasetsearch}. Thus, in addition to extracting relationships from \path{schema.org} markup, we propose a series of automatic approaches to infer dataset relationships. We focus specifically on evaluating the value of  metadata (not data); thus, we concentrate on provenance-based relationships. 

We evaluate four  methods: First, we extract relationships directly using \path{schema.org}.
Second, we develop  a set of heuristics tailored to each relationship type. Heuristics-based approaches are usually efficient to implement. Finally, we propose two machine-learning–based approaches: a classical ML approach consisting of a gradient boosted decision trees classifier and a generative AI approach using a LLM-based classifier. Each of these models represents a larger class of methods that can be used to tackle this problem setting. Section~\ref{sec:evaluation} compares the accuracy of these approaches on a large ground-truth set.

\subsection{Semantic Markup Analysis} 
We use the \path{schema.org} relationships that metadata explicitly captures: \emph{replica} (\path{schema.org/sameAs}) and \emph{derivation} (\path{schema.org/isBasedOn}). We consider datasets $A$ and $B$ to be replicas if one contains the DOI or URL of the other in its \path{sameAs} field. Dataset $A$ is considered derived from dataset $B$ if dataset $A$ contains the DOI or URL of dataset $B$ in its \path{isBasedOn} field.
 
\subsection{Heuristics-Based Methods}
We define a set of heuristics based on regularities observed by analyzing a large corpus of metadata for datasets on the Web. All comparisons in this section use normalized names and descriptions; specific rules are tailored to each relationship type. Two datasets are \textbf{replicas} if their normalized names and descriptions are exact matches or one is a non-trivial prefix of the other. Two datasets are \textbf{versions} if their normalized names are the same except for the version number, which we extract using a regular expression. Two datasets are \textbf{variants} if their normalized names are the same except for months or dates, which we extract using a regular expression. Two datasets are also variants if their prefixes before a common delimiter are the same, but suffixes are different. Here we consider two dimensions for variants: temporal and spatial.

Directional relationships require complex rules to identify. We identify \textbf{subsets} by splitting dataset names into prefixes and suffixes based on a common delimiter. Dataset $A$ is a subset of Dataset $B$ if the name prefix of $A$ and $B$ are exact matches and only Dataset $A$ has a suffix. Additionally, Dataset $A$ is a subset of Dataset $B$ if the two dataset names are the same after extracting a year or month from Dataset $A$ but not from Dataset $B$. To identify the \textbf{derived} relationship, we use observed text patterns in the corpus, such as "analysis of." Dataset $A$ is derived from Dataset $B$ if the processed names are the same after removing these patterns from Dataset $A$ but not $B$. Table~\ref{tab:rule_based_methods} shows an example of one of the heuristics used to identify the subset relationship.

\begin{table*}
\caption{An example of a heuristic to discover subsets in our corpus. We consider $D_a$ to be a subset of $D_b$}
\begin{tabular}{llll}
\hline
\textbf{ID} & \textbf{Dataset Name} & \textbf{Prefix} & \textbf{Suffix} \\
\hline
$D_a$ & "Survey of Earned Doctorates - 2019" & "Survey of Earned Doctorates" & "2019"\\
$D_b$ & "Survey of Earned Doctorates" & "Survey of Earned Doctorates" & "" \\
\hline
\end{tabular}
\label{tab:rule_based_methods}
\end{table*}

\subsection{Gradient Boosted Decision Trees Based Classification}
\label{sec:gbdt}
We used the ydf-implementation of GradientBoostedTreesLearner in TensorFlow2 \cite{ydf-trees} to train a GBDT-based multi-class classifier using manually annotated examples described in Section~\ref{sec:data}. We trained the model with a batch size of 128, a local growth method to optimize a cross entropy loss function, a random sampling method, a maximum depth of 4, a sparse oblique splitting method \cite{GBDT-2020} and a shrinkage of 0.0887 as set by a hyperparameter sweep.

\subsection{LLM-based Classification}
We fine-tuned the t5x \cite{t5x} implementation of the T5.1.1 large-language model \cite{t5} to perform a multi-class classification task using the same manually annotated training examples described in Section~\ref{sec:gbdt}. We used a batch size of 64 with 1,050,000 training steps and a learning rate of $1e-3$ set by a hyperparameter sweep.

\section{Evaluation and results}

\label{sec:evaluation}

To understand which approach works best in practice, we compared the performance of the four methods from the previous section on manually annotated ground truth data. We then apply the best-performing method to a large corpus of datasets on the Web in order to understand the prevalence of different provenance relationships between datasets.

\subsection{Training and evaluation data}
\label{sec:data}
\subsubsection{Dataset corpus.} 
We generated a corpus of dataset metadata by crawling the Web to find pages with \path{schema.org} metadata indicating that the page contains a dataset. We then selected a subset of citable datasets: we categorize a dataset as \emph{citable} if it has a persistent de-referencible identifier, like a digital object identifier (DOI). This corpus includes 2.7 million dataset-metadata entries.

\subsubsection{Ground truth.}
To generate ground truth for training and evaluation, the paper authors manually labeled 2,178 dataset pairs. The labelers had access to all metadata fields for these datasets, not just their names and descriptions. 

We observed that some relationships (e.g., replica) are much more common than others (e.g., subset). Thus, our goal when generating a set of pairs for manual labeling was to ensure that this set likely had examples of all the relationships. We used the following procedure:  
We randomly sampled 125 \emph{seed datasets} from the corpus. Each seed dataset must have between 1 and 10 replicas, as identified by a heuristic-based method (Section~\ref{sec:methods}). Note that we required that a dataset \emph{has} a replica but not that the replica is included in the sample. We observed that datasets with replicas are also likely to have other related datasets. Seed datasets also had to come from hosts with more than 30 datasets. We limited seed datasets  to a maximum of 2 datasets per host to ensure diversity of the set. 
We projected the metadata of the datasets in our sample and our corpus into the NewsEmbed~\cite{liu-newsembed} embedding space (an embedding space that we found worked particularly well for datasets). Given a seed dataset $S$ and its 20 nearest neighbors $\mathcal{D} = \{D_1,...,D_{20}\}$, we label manually  each relationship $\{\langle S, D_1\rangle,\ldots, \langle S, D_{20} \rangle\}$ and a random sample of 20 relationships between $\langle D_x, D_y \rangle$ where $D_x \in \mathcal{D}$ and $D_y \in \mathcal{D}$. 

The number of subset and derived relationships in a random sample was still very low. Typical data interpolation techniques were not possible because subset and derived relationships are non-reflexive. To address the sparse label space, we added government datasets used in Show US the Data Kaggle Competition~\cite{Lane2022Data}. We observed that these datasets were more likely to have subsets or derivations.

For machine-learning based methods, we used 70:15:15 split for training, validation, and evaluation data. For other methods, we used the same evaluation data as for the ML-based methods (the 15\% of the labeled pairs).

\subsection{Results}
\label{sec:results}

Table~\ref{tab:results} presents the comparison of evaluating the four methods from Section~\ref{sec:methods}. Our experiments find that \path{schema.org} metadata alone is insufficient for identifying relationships between datasets, even for the two types for which \path{schema.org} exist (replica and derived): Indeed, \emph{no} pairs of datasets in our random sample had an explicit relationship defined between them.

% Table~\ref{tab:results} presents the comparison of evaluating the four methods described in Section~\ref{sec:methods}. Our experiments find that \path{schema.org} metadata from providers alone is insufficient for identifying relationships between datasets even for the two types for which \path{schema.org} exist (replica and derived): Indeed, \emph{no} pairs of datasets in our random sample had an explicit relationship defined between them. 

\begin{table*}
\centering
\caption{Precision (P), recall (R), and F1 scores for each method and relationship type. The scores for the method with the top F1 score for each relationship is bolded.}
\begin{tabularx}{\textwidth}{l c *{12}{Y}}
\toprule
\multirow{2}{*}{} &
  \multicolumn{3}{c}{\textbf{Schema.org}} &
  \multicolumn{3}{c}{\textbf{Heuristics}} &
  \multicolumn{3}{c}{\textbf{GBDT}} &
  \multicolumn{3}{c}{\textbf{T5}} \\
\cmidrule(lr){2-4} \cmidrule(lr){5-7} \cmidrule(lr){8-10} \cmidrule(lr){11-13}
\textbf{Relationship} & \textbf{P} & \textbf{R} & \textbf{F1} & \textbf{P} & \textbf{R} & \textbf{F1} & \textbf{P} & \textbf{R} & \textbf{F1} & \textbf{P} & \textbf{R} & \textbf{F1} \\ 
\midrule
Replica & 0.00 & 0.00 & 0.00 & 1.00 & 0.35 & 0.51 & \textbf{0.97} & \textbf{0.95} & \textbf{0.96} & 0.92 & 0.92 & 0.92  \\
Version & N/A & N/A & N/A & 0.96 & 0.53 & 0.68 & 0.92 & 0.80 & 0.86 & \textbf{0.87} & \textbf{0.87} & \textbf{0.87}  \\
Subset & N/A & N/A & N/A & 0.49 & 1.00 & 0.65 & \textbf{1.00} & \textbf{0.80} & \textbf{0.89} & 0.82 & 0.90 & 0.86  \\
Derived & 0.00 & 0.00 & 0.00 & 0.00 & 0.00 & 0.00 & 0.67 & 0.33 & 0.44 & \textbf{0.80} & \textbf{0.67} & \textbf{0.73}  \\
Variant & N/A & N/A & N/A & 1.00 & 0.50 & 0.67 & \textbf{0.90} & \textbf{0.85} & \textbf{0.87} & \textbf{0.81} & \textbf{0.93} & \textbf{0.87}  \\
None &  0.33 & 1.00 & 0.49 & \textbf{1.00} & \textbf{0.80} & \textbf{0.89} & \textbf{0.85} & \textbf{0.93} & \textbf{0.89} & \textbf{0.94} & \textbf{0.85} & \textbf{0.89}  \\
    \bottomrule
    \end{tabularx}
\label{tab:results}
\end{table*}

Heuristics-based methods perform reasonably well for certain relationship types, such as \emph{none} (i.e., there is no relationship between two datasets) and \emph{version}. These methods often have low recall because they are brittle: small perturbations in names or descriptions of datasets significantly affect their performance. The GBDT Classifier and T5-Based Classifier both perform quite well and have similar F1 scores for all relationships except \emph{derived}. For the \emph{derived} relationship, which is more semantically complex, the T5-Based classifier outperforms the GBDT Classifier. 

\begin{figure}
\centering
\caption{The overall accuracy for each method type.}
\includegraphics[width=8cm]{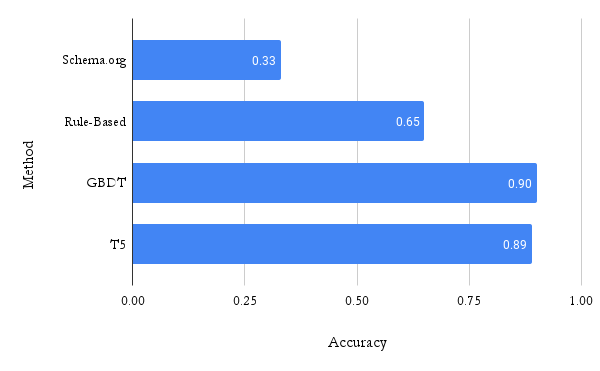}
\label{fig:accuracy}
\end{figure}

Figure~\ref{fig:accuracy} compares the overall accuracy of the methods. We opt to calculate accuracy as opposed to macro-average precision and recall because there exists a large imbalance between the prevalence of each relationship on the Web.

\subsection{Corpus-Level Analysis}
We analyzed the corpus of 2.7 million citable datasets (Section~\ref{sec:data}) using the GBDT classifier because it had the highest overall accuracy. Because classifying all of the $O(N^2)$ pairs of datasets is extremely  expensive computationally, we clustered datasets first and then classified pairs within each cluster. Specifically, we used the NewsEmbed embedding space and classified a relationship between each dataset and its 20 nearest neighbors. This process gave us 42.4 million unique dataset pairs for the GBDT classifier to classify. 

Out of 2.7M datasets in the corpus, 20.1\% had at least one relationship with another dataset, and 22\% had multiple relationships. Table~\ref{tab:corpus_level_results} shows the distribution of the identified relationships, with the \emph{replica} relation being the most prevalent. Only a handful of datasets had \emph{derived} or \emph{version} relation.

\begin{table}
\centering
\caption{The distribution of relationship in the corpus.}
\begin{tabular}{ll}
\hline
Relationship Type & Percentage \\
\hline
Replica & 77.8\% \\
Subset & 14.2\% \\
Variant & 5.2\% \\
Derived & 1.8\% \\
Version & 1.0\% \\
\hline
\end{tabular}
\label{tab:corpus_level_results}
\end{table}

Datasets in each replica pair come from different sites by definition. In order to understand the publishing ecosystem better, we looked at whether or not pairs of datasets in other relationships come from the same repository or tend to be distributed. 
Out of 61,485 dataset subsets, 59\% are from the same site as the parent dataset; the rest are from different sites. Conversely, the majority of variants, derived datasets, and versions exist on the same site: 79\% of variants, 83\% of derived datasets, and 97\% of versions.

\section{Discussion and future work}
\label{sec:discussion}

Our categorization of dataset relationships and corpus analysis highlighted both the complexity of and the need for identifying these relationships.

\subsection{Relationships are, indeed, complicated}

Several previous works categorized dataset relationships (see Section~\ref{sec:related-work}).
Our categorization in Section~\ref{sec:dataset-relationships} complements this effort. Critically, the user-centric approach gave us a unique view. Consider, for example, the \emph{subset} and the \emph{integrated} relationships. In both cases, datasets share some of the content: with the subset relationship, they share some or parts of records. For datasets to be joinable, they must share a subset of foreign keys. However, if we approach the distinction from a user point of view, these two relationships are quite different. Usually we seek \emph{subset} relationships for a manageable dataset slice, whereas for joinable datasets, we aim to expand existing data for new insights.

The user-centric view also helps us decide whether to include specific relationships in our taxonomy. For instance, we have not found the \emph{replica} relationship in other categorizations. However, in the context of Web-based dataset discovery, users need to understand the relationships between original sources of datasets and repositories that aggregate datasets from multiple sources. Users then can choose a site that they find reliable and trustworthy to download the dataset.

Having the grounding in user tasks did not eliminate the need for difficult decisions. For example, determining topic similarity between datasets raises questions. While two datasets covering different weather aspects in the same region for the same timespan are clearly related, the situation differs with education outcome datasets. If one covers high schools and the other elementary schools, they may be considered topically similar only if the user's interest spans all school education. Topic similarity varies based on user task granularity.

In our own work, we implemented a dataset-discovery tool based on the Web-based corpus that we described. We use several of the relationships from Section~\ref{sec:dataset-relationships} directly in the tool. In dataset results, we group together \emph{replicas} of a dataset giving the user an option to get the dataset from their preferred site. We also group \emph{versions} and \emph{variants} in order to simplify navigation and show the larger diversity of search results. In the current draft, we omit references to the tool and screenshots to preserve the anonymity of the submission.

\subsection{Semantic markup for dataset relationships}

Our analysis found that over 20\% of datasets have at least one relationship with another dataset. These relationships are not captured by \path{schema.org} metadata. While some researchers noted the low quality of semantic markup (e.g., \cite{alrashed-datasetness}), we found it to be extremely incomplete. If \path{schema.org/sameAs} relationships were accurate and complete, our replica-identification mechanism would not be needed. The metadata connecting datasets is often incomplete or inaccurate for several reasons. First, the community can improve and expand \path{schema.org/Dataset} properties, as some definitions are vague. For example, \texttt{isBasedOn} can capture subset, revision, or version relationships. \path{Schema.org} also lacks relationships like linking to a previous version. Second, our analysis, supported by \cite{Klump-2021,herschel2017survey,zuiderwijk2020drives}, highlights the need for data sharing best practices, including publishing datasets with digital object identifiers, linking datasets to papers, and capturing provenance information in metadata. Additionally, the lack of tools utilizing metadata may deter authors from providing accurate metadata; developing such tools could incentivize authors to improve metadata accuracy.

\subsection{Limitations in future work}
We analyzed millions of datasets, optimizing the process to find pairs of datasets to label. The ground truth datasets were limited to pairs that are neighbors in an embedding space, focusing on datasets with similar names and descriptions. Our methods may miss relationships when dataset names change significantly, although minor changes like adding acronyms should not be affected. 

We used a few metadata fields to infer relationships, with future analysis planned to explore the impact of fields like authors, providers, and explicit temporal or spatial coverage values.

We derived our relationships using the corpus of datasets on the Web. These relationships likely paint a different picture for datasets in specific dataset repositories (e.g., Figshare, Zenodo), and also among the datasets that constitute the linked open data (LOD) cloud. Understanding how these relationships apply to the datasets in LOD will enable us to highlight similarities and differences between linked data and more traditional data.

Identifying provenance-based relationships lays the groundwork to studying data quality and trustworthiness of data changes on the Web, aiding users in finding reliable data sources and identifying information gaps.

Finally, understanding non-provenance relationships is the first step in helping users find the right data for their tasks. Users often seek data to complete specific tasks (e.g., training an ML model for weather prediction). Understanding dataset relationships helps us better assist users in finding the necessary data.

\section{Conclusion}

Understanding relationships between datasets is crucial for extracting valuable insights that can drive innovation and positively impact various domains. Using even simple analysis methods, we can see that datasets on the Web are connected in many different ways. However, while this analysis can help in identifying some of the relationships, research communities must develop best practices that encourage dataset authors to specify metadata. Overall, our paper sets a benchmark for future research and highlights the importance of understanding dataset relationships for scientific research and decision-making processes.

\paragraph*{Supplemental Material Statement:} \textit{We publish metadata and dataset relationships for the collection of 2.7 million dataset pages analyzed in Section 6 on Figshare (see Footnote 1). 
While the source code for the methods outlined in Section 5 references proprietary libraries and therefore cannot be released, we have reproduced the overall code logic in pseudocode, also available at the aforementioned link. The base t5x implementation described in Section 5.3 can be found on GitHub.\footnote{\url{https://github.com/google-research/t5x}} The GBDT training algorithm described in Section 5.4 can be found on GitHub.\footnote{\url{https://github.com/google/yggdrasil-decision-forests}}}

\bibliographystyle{splncs04}
\bibliography{references}

\end{document}